\documentclass[10pt,superscriptaddress,twocolumn,amsmath,amssymb,aps,prl,showpacs]{revtex4}
\usepackage{mathrsfs}
\usepackage{graphicx}
\usepackage{lipsum}
\usepackage{dcolumn}
\usepackage{bm}
\usepackage{amssymb}
\usepackage{amsmath}
\usepackage{paralist}
\usepackage[colorlinks,linkcolor=red,anchorcolor=blue,citecolor=blue]{hyperref}

\usepackage{pdfpages}
\usepackage{subfigure}
\usepackage{cleveref}

\allowdisplaybreaks[4]

\newcommand{\gp}{Gr\"uneisen parameter\,}
\newcommand{\gps}{Gr\"uneisen parameters\,}
\newcommand{\mr}{\mathrm{d}}
\newcommand{\me}{\mathrm{e}}
\newcommand{\Li}{\mathrm{Li}}

\newcommand{\tA}{\tilde{A}}
\newcommand{\tAu}{\tilde{A}^{(1)}}
\newcommand{\tAb}{\tilde{A}^{(2)}}
\newcommand{\tFu}{F^{(1)}}
\newcommand{\tFb}{F^{(2)}}
\newcommand{\tFut}{F^{(1)}_\frac{3}{2}}
\newcommand{\tFbt}{F^{(2)}_\frac{3}{2}}
\newcommand{\tFuo}{F^{(1)}_\frac{1}{2}}
\newcommand{\tFbo}{F^{(2)}_\frac{1}{2}}
\newcommand{\tFumo}{F^{(1)}_{-\frac{1}{2}}}
\newcommand{\tFbmo}{F^{(2)}_{-\frac{1}{2}}}

\newif\ifNOSUP \NOSUPfalse

\begin{document}

\title{Gr\"{u}neisen Parameters:  origin, identity and quantum refrigeration }

\author{Yi-Cong Yu}
\affiliation{State Key Laboratory of Magnetic Resonance and Atomic and Molecular Physics,
Wuhan Institute of Physics and Mathematics, IAPMST,  Chinese Academy of Sciences, Wuhan 430071, China}

\author{Shizhong Zhang}      
   \email[]{shizhong@hku.hk}
\affiliation{Department of Physics and Center of Theoretical and Computational Physics, The University of Hong Kong, Hong Kong, China}

\author{Xi-Wen Guan}
\email[e-mail:]{xwe105@wipm.ac.cn}
\affiliation{State Key Laboratory of Magnetic Resonance and Atomic and Molecular Physics,
Wuhan Institute of Physics and Mathematics, IAPMST, Chinese Academy of Sciences, Wuhan 430071, China}
\affiliation{Department of Theoretical Physics, Research School of Physics and Engineering,
Australian National University, Canberra ACT 0200, Australia}

\begin{abstract}

In solid state physics, the Gr\"{u}neisen parameter (GP), originally introduced in the study of the effect of changing the volume of a crystal lattice on its vibrational frequency, has been widely used to investigate  the characteristic energy scales  of systems with respect to  the changes of  external  potentials.  On the other hand, the  GP is  little investigated in a strongly interacting quantum gas systems. Here we report  on our general results on the origin of GP, new identity and caloric effects in quantum gases of ultracold atoms. We prove that the symmetry of the dilute quantum  gas systems leads to a simple identity among three different types of GPs, quantifying caloric effect induced respectively by variations of volume, magnetic field and interaction. Using exact Bethe ansatz solutions, we present a rigorous study of these different GPs and the  quantum refrigeration  in one-dimensional  Bose and Femi gases. Based on the exact equations of states of these systems, we  obtain analytic results for the singular behaviour of the GPs and the caloric effects at quantum criticality. We also predict the existence of the lowest temperature for cooling  near a quantum  phase transition. It turns out that the interaction ramp-up and -down in quantum gases  provides a promising protocol of quantum refrigeration   in addition to the usual adiabatic demagnetization cooling in solid state materials.

\end{abstract}

\maketitle

\section{I. Introduction}       
The structure of energy spectrum of a quantum many-body system and its evolution under external perturbation characterises essentially its possible phases. As an example, the \gp (GP)~\cite{Gruneisen_AdP_1908,Gruneisen_AdP_1912}, which was introduced by Eduard Gr\"uneisen in the beginning of 20th Century in the study of the effect of volume change of a crystal lattice on its vibrational frequencies,  has been extensively studied for the exploration of caloric effect of solids and phase transitions associated with volume change.  Similarly, the magnetic GP quantifies the magnetocaloric effect (MCE), establishing connection between refrigeration  and variation of magnetic field.

So far, the GP has found diverse applications in geophysics \cite{Stacey_PEPI_1995,Shanker_PEPI_2017}, chemical physics\cite{Mausbach_JCP_2016, Liu_PCCP_2017}, high pressure physics and plasma physics \cite{Wang_PRE_2017,Kumar_PJP_2016,Khrapak_PP_2017}. Recently, experiments also started to focus on GP in heavy-fermion systems \cite{Kuchler_PRL_2003,Kuchler_PRL_2006,Kuchler_STAM_2007}, in which the physical properties at low temperatures are dominated by $f$-electrons and their antiferromagnetic exchange $J$ with conduction electrons \cite{Gegenwart_RPP_2016}.  The heavy-fermion metals are extremely sensitive to a small change of pressure and this pressure sensitivity is reflected in highly enhanced values of the GP \cite{Steppke_Science_2013}. At low temperatures, divergence of the GP stronger than logarithmic upon cooling in the quantum regime is used for experimental identification of quantum critical points \cite{Kuchler_PRL_2003,Tokiwa_PRL_2009,Wolf_PNAS_2011,Wolf_IJMPB_2014,Ryll_PRB_2014}.
 
 The corresponding GPs in dilute quantum gases, however, are much less explored. In addition to the ability to change the volume of the system by modifying the external confining potential and the ability to change the equivalent magnetic field by changing population imbalance, it is also possible to change the interaction directly by using Feshbach resonance \cite{Chin_RMP_2010,Guan_RMP_2013}. This possibility suggests a new avenue for studying a novel interacting GP in addition to those defined by changes in volume and magnetic field~\cite{Zhu_PRL_2003,Yoshinori_JPCM_2006,Kuchler_PRL_2006,Kuchler_STAM_2007,Tokiwa_PRL_2009,Wolf_PNAS_2011,Smith_AEM_2012,Steppke_Science_2013,Ryll_PRB_2014,Kumar_PJP_2016,Gegenwart_RPP_2016,Mousumi_JPD_2016,Breunig_ScienceAdvance_2017,Liu_SSC_2017} . Furthermore, we establish an exact identity between these various GPs, making use of the scaling properties of the quantum gas system. The interaction driven caloric effect in quantum gases  will also be discussed.

In this paper, we discuss the physical origin of GPs, establish a new identity and  investigate the efficiency of quantum refrigeration by modifying the interaction strength in  quantum gases. These GPs can be generally described by the ratio between the energy-pressure (or energy-magnetization and energy-contact) covariance and energy fluctuation, similar to the the Wilson ratio \cite{Wilson_1975}. Using exact Bethe ansatz solutions, we show how interacting GP  characterizes the quantum phase transitions for 1D quantum gases and demonstrate that  the cooling effect is greatly enhanced near the quantum critical point. Our study shows promising  route for studying interaction driven quantum heat engine and  refrigeration in ultracold atoms. 


\section{II. Theory:  the origin, generalization and new identity    } 

{\bf 1.  The origin. }
There are many formulations of the GP to quantify the degree of anharmonicity on the structure of the energy spectrum in response to volume change.  The original definition of the GP was introduced by E. Gr{\"u}neisen for  the  Einstein model \cite{Gruneisen_AdP_1908,Gruneisen_AdP_1912}, 
\begin{align}
\Gamma =: -\frac{V}{\omega_0}\frac{\partial \omega_0}{\partial V} = \frac{V}{C_V}\frac{\partial S}{\partial V},
\label{eq:definition_Gruneisen}
\end{align}
where the  excitations in a solid is described by $N$ phonons with the same frequency $\omega_0$. $S$ is the entropy and $V$ denotes the volume. In quantum  statistical  physics,  the differential forms  of the internal  energy $E$ and the pressure $p$ can be represented by the fluctuations and covariances of thermodynamic quantities. If we  regard  the population $a_i$ of the 
$i$-th energy level as a  distribution function of a random variable and observable thermal quantities as the expectation value with respect to this distribution,   then one can obtain the  following differential relations, see supplementary material (SM)  \cite{Sup}
\begin{align}
& \mr E=[-\text{Cov}(E,E)]\mr \beta + [-p+\beta \text{Cov}(p,E))] \mr V\notag \\
& \mr p=[ \text{Cov}(p,E))] \mr \beta+[E^{\prime \prime}+\beta \text{Cov}(p,p)]\mr V,
\notag 
\end{align}
where $\text{Cov}$ denotes the covariance and $E^{\prime \prime}=: \sum_i a_i \frac{\partial^2 \epsilon_i}{\partial V^2} $. $\beta=1/(k_BT)$ and $k_B$ is the Boltzman constant and $T$ is the temperature. Then the GP  is simply given by 
\begin{align}
\Gamma=-\frac{V \text{Cov}(p,E)}{\text{Cov}(E,E)}=\frac{Vdp/d\beta|_V}{dE/d\beta|_V}. 
\label{eq:fluctuation_relation}
\end{align}
Thus, in this case, $\Gamma$ represents the relative importance of energy-pressure covariance and the energy fluctuation in the system.  In contrast to the susceptibility (or compressibility) Wilson ratio proposed in  \cite{Wilson_1975,Yu_PRB_2016,He_PRB_2017}, i.e., the ratio between the {magnetization $M$ (or particle number) fluctuation and the energy fluctuation, namely $R_W^{\chi} \propto \frac{\text{Cov}(M,M)}{\text{Cov}(E,E)}$ (or $R_W^{\kappa} \propto \frac{\text{Cov}(N,N)}{\text{Cov}(E,E)}$),  the GP (\ref{eq:fluctuation_relation})  provides additional insights into the spectral information with respect to  the change of the volume of the system. 

{\bf 2. In grand canonical ensemble.} In cold atom system, it is far more convenient to work in grand canonical ensemble and it is useful to derive the form of $\Gamma$ in the grand canonical ensemble. Let $\mu$ be the chemical potential of the system and one finds
\begin{equation}
\Gamma=\frac{V\left.\frac{\mathrm{d}p}{\mathrm{d}T}\right|_{V,N}}{\left.\frac{\mathrm{d}E}{\mathrm{d}T}\right|_{V,N}}
=\frac{1}{T} 
\frac{\frac{\partial^2 p}{\partial \mu^2}\frac{\partial p}{\partial T}
-\frac{\partial^2 p}{\partial \mu \partial T}\frac{\partial p}{\partial \mu}}
{\frac{\partial^2 p}{\partial \mu^2}\frac{\partial^2 p}{\partial T^2}
-(\frac{\partial^2 p}{\partial \mu \partial T})^2},
\label{eq:calculation_for_Gamma}
\end{equation} 
In deriving the above equations,  we applied the Maxwell's relations and  used  the homogeneous assumption that the grand thermal potential is a linear function of the volume neglecting the surface effect in the thermodynamic limit \cite{Landau_1980e}, i.e. $\Omega = -pV$ \footnote{Here we do not write out explicitly  the vertical lines and corresponding variables in the partial derivatives, where 
the derivatives are calculated in the grand canonical ensemble with variables $V$(volume),$T$(temperature),$H
$(magnetic filed),and $c$(coupling strength).  }.

 {\bf 3.  The effective and magnetic \gp}. There is  a widely  used    effective GP in experiment, defined as    the ratio of thermal expansion parameter $\beta_T=\frac{1}{V}\frac{\partial V}{\partial T}\mid_{p,N}$ to the specific heat at a constant volume 
\cite{Zhu_PRL_2003,Kuchler_PRL_2006,Kuchler_STAM_2007,Weickert_PRB_2012,Steppke_Science_2013,Liu_SSC_2017} 
\begin{align}
\Gamma_{\text{eff}}=\frac{\beta_T}{c_V/V}=\Gamma \cdot
\frac{\partial^2 p}{\partial \mu^2} 
\left(\frac{\partial p}{\partial \mu}\right)^{-2}=\Gamma \cdot \frac{\kappa }{n^2},
\end{align}
 here $\kappa$ is the compressibility and $n$ is the density. 
We denote it by "eff-\gp " since it is not equivalent to the original definition  (\ref{eq:calculation_for_Gamma}).  In the above equation,  the thermal expansion parameter in grand canonical ensemble  can be given by 
\begin{eqnarray}
\beta_T=\left(\frac{\partial^2 p}{\partial \mu^2}\frac{\partial p}{\partial T}
-\frac{\partial^2 p}{\partial \mu \partial T}\frac{\partial p}{\partial \mu}
\right)\left(\frac{\partial p}{\partial \mu}\right)^{-2}.
\end{eqnarray}
Note that  the usefulness of eff-GP is not well established in experiment; see the discussion on its divergent  behaviour at quantum critical points \cite{Breunig_ScienceAdvance_2017,Steppke_Science_2013,Kuchler_STAM_2007}.
However, it is clear that the eff-GP is not a dimensionless parameter and shows different scaling forms at the quantum critical points. To clearly show the dimensionless nature of the \gp,   we define another dimensionless GP by \cite{Sup}
\begin{align}
\Gamma=\frac{V\frac{\partial S}{\partial V}\mid_{N,T}}
{T\frac{\partial S}{\partial T}\mid_{N,V}}
\label{eq:form_definition}
\end{align}
 which is equivalent to the definition (\ref{eq:definition_Gruneisen}), (\ref{eq:fluctuation_relation}) and (\ref{eq:calculation_for_Gamma}) and is intimately related to the expansionary caloric effect
\begin{eqnarray}
 \left.\frac{\partial T}{\partial V}\right|_{S,N,H}&=&\frac{T}{V} \Gamma. \label{GP-1}
\end{eqnarray}
In quantum statistics,  the volume $V$ of a system can be regarded as an external filed that  imposes a constrain on the particles. Therefore, it is natural  to investigate other  potentials that impose different constraints  on the system.  As a remarkable example, the well known magnetic GP discussed in experiments \cite{Tokiwa_PRL_2009,Wolf_PNAS_2011,Steppke_Science_2013,Ryll_PRB_2014} can be introduced analogously by replacing 
the volume $V$ by the magnetic field $H$ in the definition (\ref{eq:form_definition})
\begin{align}
\Gamma_{\text{mag}}=-\frac{H\frac{\partial S}{\partial H}\mid_{N,T,V}}
{T\frac{\partial S}{\partial T}\mid_{N,B,V}}
\label{eq:definition_mag_gruneisen}
\end{align}
Here we added  a minus sign following the former work  \cite{Garst_PRB_2005,Ryll_PRB_2014},
and put the magnetic field $H$ in the numerator in order to  keep the GP  dimensionless.
It is straightforward to  obtain  the explicit form of the magnetic GP in grand canonical ensemble 
\begin{align}
\Gamma_{\text{mag}}=-\frac{H}{T} \frac{\frac{\partial^2 p}{\partial \mu^2}\frac{\partial^2 p}{\partial H \partial T}
-\frac{\partial^2 p}{\partial \mu \partial H}\frac{\partial^2 p}{\partial \mu \partial T}}{\frac{\partial^2 p}{\partial \mu^2}\frac{\partial^2 p}{\partial T^2}
-(\frac{\partial^2 p}{\partial \mu \partial T})^2}.
\label{eq:calculate_mag_gamma}
\end{align}
The magnetic GP  (\ref{eq:definition_mag_gruneisen})   plays an important  roles in  the experimental study  of  solid state materials \cite{Gegenwart_RPP_2016,StraBel_PRB_2015,Mousumi_JPD_2016,Ryll_PRB_2014,Weickert_PRB_2012}. 

One of the most important features of the magnetic materials is the magnetocaloric effect, related to the  magnetocaloric refrigeration. 
 In low temperature physics,  this  is known as adiabatic demagnetization cooling; see recent new developments \cite{Wolf_PNAS_2011,
Wolf_IJMPB_2014}.
By the definition of the $\Gamma_{\text{mag}}$ in \cref{eq:definition_mag_gruneisen}, we further  get 
\begin{align}
\left.\frac{\partial T}{\partial H} \right|_{S,N,V}=\frac{T}{H} \Gamma_{\text{mag}},
\label{eq:MCE}
\end{align}
which establishes an important relation between magnetocaloric effect and the magnetic GP.  Experimentally, it is easier to measure the magnetocaloric effect and from Eq. (\ref{eq:MCE}), obtain $\Gamma_{\text{mag}}$ instead of using its original definition (\ref{eq:definition_mag_gruneisen}). One can obtain the 
magnetic entropy change $\partial S/\partial H\mid_{N,T,V}$ once we know the value of specific heat. The magnetic GP contains information free of any material-specific parameter  \cite{Ryll_PRB_2014}.

%

{\bf 4.  The interacting  \gp.} In addition to the usual conjugate variables that one usually encounters in thermodynamics, in ultracold atomic gases, it is also possible to define another set of conjugate variable related to the interaction between atoms. In the case of $s$-wave interacting system, the low-energy scattering properties are determined entirely by the $s$-wave scattering length $a_s$. In one-dimensional system, the 1D coupling constant $c$ is related to the scattering length ($c\propto a_s^{-1}$). In reality, it is possible to change the scattering length $a_s$ by using Feshbach resonance and one can define analogously another GP related to interaction
\begin{align}
\Gamma_{\text{int}}=-\frac{c\frac{\partial S}{\partial c}\mid_{N,H,T,V}}
{T\frac{\partial S}{\partial T}\mid_{N,H,c,V}}
=-\frac{\frac{\partial^2 p}{\partial \mu^2}\frac{\partial^2 p}{\partial c \partial T}
-\frac{\partial^2 p}{\partial \mu \partial c}\frac{\partial^2 p}{\partial \mu \partial T}}{\frac{\partial^2 p}{\partial \mu^2}\frac{\partial^2 p}{\partial T^2}
-(\frac{\partial^2 p}{\partial \mu \partial T})^2}
\frac{c}{T}. 
\label{eq:definition_coupling_gruneisen}
\end{align} 
The physical significance of $\Gamma_{\text{int}}$ is that it describes the caloric effect due to modification of interaction strength. In particular, in an isentropic process, one can relate the change of temperature to interaction strength given by 
\begin{equation}
\left.\frac{\partial T}{\partial c} \right|_{S,N,V,H}=\frac{T}{c} \Gamma_{\text{int}}.
\label{eq:CMCE}
\end{equation}
This is an interaction  analog of  the magnetocaloric effect. We observe that from \cref{eq:CMCE} that 
a heat engine and quantum refrigeration can be constructed by tuning the interaction strength in quantum gases. 
Therefore the interaction gradients are capable of cooling the system of interacting fermions like the magnetization gradient cooling \cite{Medley-PhysRevLett.106.195301,Weld-PhysRevA.82.051603}.
 We shall further discuss interaction driven quantum refrigeration in  next section.

{\bf 5.  Universal identity.} So far we have  presented   three  different GPs, i.e., $\Gamma,\,\Gamma_{\text{mag}}$ 
and $\Gamma_{\text{int}}$, which quantify  the  degrees of anharmonicity of spectral  structures in regard  of the variations of volume, magnetic field and interaction strength, respectively. 
Using the general thermal potential \cite{Yang_NC_2014}, one can find a new identity for the three GPs for dilute system described by $s$-wave scattering length $a_s$ \cite{Chin_RMP_2010}. For these systems, one has the following scaling transformations:
$L \rightarrow \me^{\lambda} L,  c \rightarrow \me^{\chi\lambda}c, 
H \rightarrow \me^{-2\lambda}H$, where $\me^{\lambda}$ is the scaling 
amplitute  and the $\chi$ is the exponent of the dependency of the coupling strength and the scattering length by $c \propto a_s^\chi$, then we  find that the spectrum of such  quantum many-body systems will be changed by
$\epsilon_n \rightarrow \me^{-2\lambda} \epsilon_n$, here $\epsilon_n$ denotes the $n$-th energy level. Meanwhile, if the temperature transforms as $T \rightarrow \me^{-2\lambda}T$, the population $a_n = \me^{-\frac{\epsilon_n}{T}}/\sum_i \me^{-\frac{\epsilon_i}{T}}$ is invariant under such  scaling transformations, 
and so does the entropy $S = -\sum a_i \ln a_i$, i.e.
%
%
%
\begin{eqnarray}
0 = \mr S &=& \left.\frac{\partial S}{\partial V}  \right|_{\small T,H,c} \mr V
+ \left.\frac{\partial S}{\partial T}\right|_{\small V,H,c} \mr T\nonumber\\
&&
+ \left.\frac{\partial S}{\partial H}\right|_{\small V,T,c} \mr H
+ \left.\frac{\partial S}{\partial c}\right|_{\small V,T,H}\mr c.\nonumber
\end{eqnarray}
Substituting the scaling transformations into the above equation and noticing 
 $V = L^d$ with $d$ being the dimension of the system, we obtain an important identity 
\begin{eqnarray}
d V\left.\frac{\partial S}{\partial V}\right|_{\small T,H,c} = 2T\left.\frac{\partial S}{\partial T}\right|_{\small V,H,c}+2H\left.\frac{\partial S}{\partial H}\right|_{\small V,T,c}
 - \chi c\left.\frac{\partial S}{\partial c}\right|_{\small V,T,H} \nonumber
\end{eqnarray}
that relates the entropy changes to the variations of the interaction, magnetic field and the volume of the system. 
Using the definitions of GPs given in \cref{eq:definition_Gruneisen,eq:definition_mag_gruneisen,eq:definition_coupling_gruneisen}, 
we obtain a simple  identity 
\begin{align}
d \Gamma +2\Gamma_{\text{mag}} - \chi \Gamma_{\text{int}}=2. 
\label{eq:identity}
\end{align}
 In one dimension systems we have $d = 1$ and $\chi = -1$ \cite{Olshanii_PRL_1998}, the identity above is reduced to $\Gamma +2\Gamma_{\text{mag}} + \Gamma_{\text{int}}=2$.
Although we obtained  this identity through the scaling invariance of the entropy, 
it is universal and valid for quantum gases in 1D and higher dimensions. 
 We can prove this  identity (\ref{eq:identity})  in a more  conventional  way of  quantum statistical physics, see \cite{Sup}. 
 This  identity \cref{eq:identity} has  many interesting applications. 
 The term $2/d$ in \cref{eq:identity} gives the nature of  the  scaling invariant  spectrum of   ideal gases, also see  
the discussion on  the \gp, where  it is  exactly $\frac{2}{3}$ for 3D free gas \cite{Souza_EJP_2016}, 
 obviously corresponding to a special case of  \cref{eq:identity} with  $d=3$ and  $\Gamma_{\text{mag}}=\Gamma_{\text{int}}=0$.
 A further study of the identity  (\ref{eq:identity}) will be published elsewhere \footnote{L. Peng,  Y.-C. Yu and X.-W. Guan, in preparation, 2019.}.

\section{III. Applications: quantum criticality and quantum refrigeration}
{\bf 1. Magnetic and interaction driven refrigeration.} Through the study presented in last section, we have shown that it is possible to reduce the temperature of a magnetic system by changing the external magnetic field in an isentropic process. Based on this magnetocaloric effect, it is possible to cool the systems into extremely low temperatures via either spin flipping or magnetic field gradient (spin transport) \cite{Medley-PhysRevLett.106.195301,Weld-PhysRevA.82.051603}. Other refrigerators are also discussed~\cite{Ekkes_JPD_2005,Khovaylo_PSS_2014}. 

%

\begin{figure*}
\includegraphics[width=0.9\textwidth]{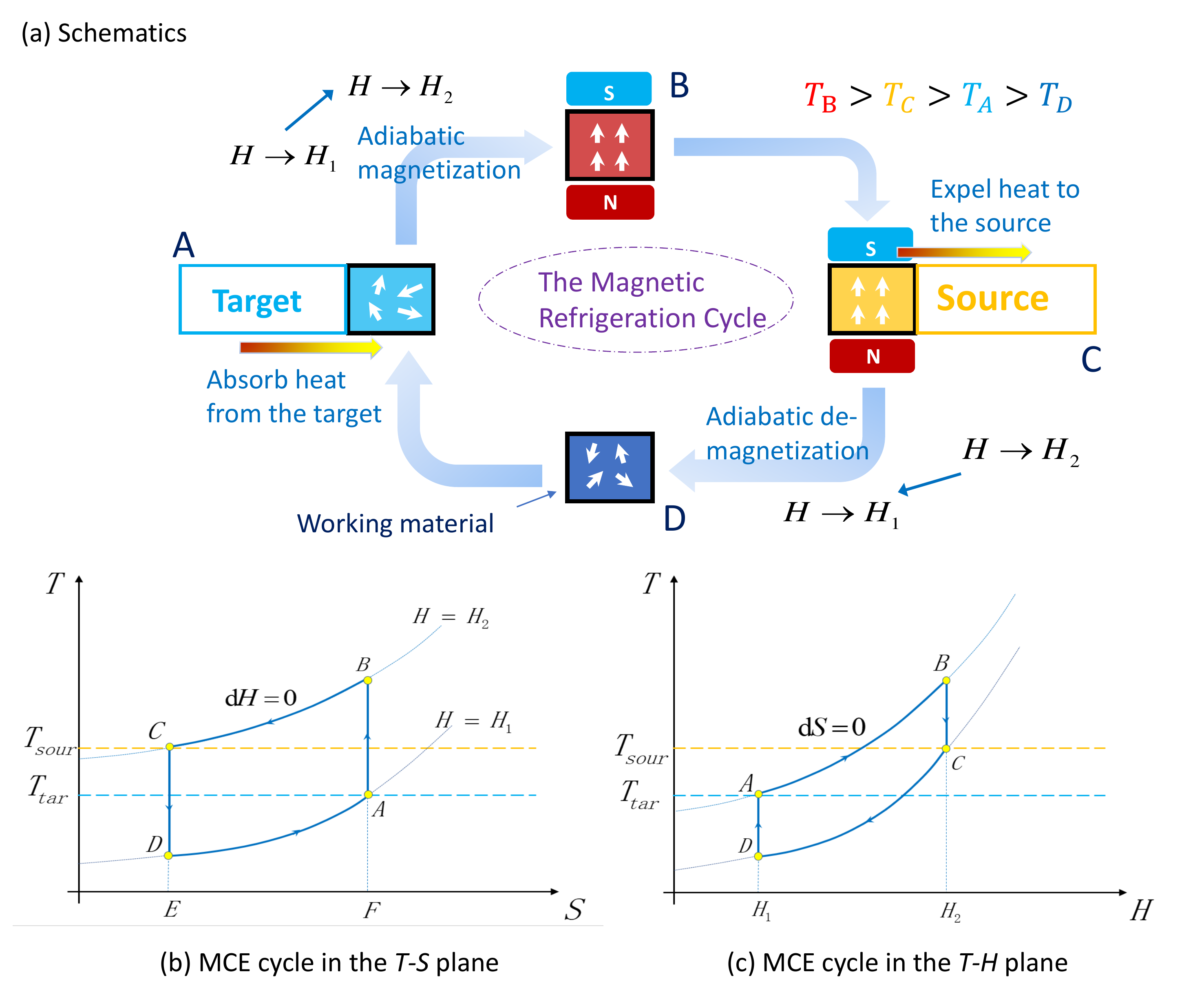}
\caption{Schematic representation for the magnetic refrigeration cycle:  (a) The working  circle between the target and the heat source absorbs heat from the target at  a lower temperature $T_{\text{tar}}$  and transfers heat  to the source at a higher temperature $T_{\text{sour}}$. 
Different colors label different temperatures.  At the point B and C,  a strong magnetic field is applied and the material is polarized sufficiently. Whereas  at the point A and D,  the material is demagnetized. 
(b) The magnetic refrigeration cycle in the $T-S$ plane and (c) in the $T-H$ plane. $B \to C$ and $D \to A$ are isomagnetic processes with $\mr H=0$, and $A \to B$ and $C \to D$ are isoentropic.  The whole processes $A \to B \to C \to D \to A$
were discussed in detail the main  text. 
In $D \to A$, the heat is absorbed by the working material from the target.
In the process $B \to C$, the heat is expelled from the working material to the heat source. }
\label{fig:schematic_MCE}
\end{figure*}

In cold atomic gas systems, however, inter-conversion between different spin (hyperfine) states is very slow and the usual magnetic cooling is inefficient. In addition, the corresponding external magnetic field is given by the difference between chemical potentials of two spin states, which is not possible to alter in experiments.  On the other hand, it is possible to construct a corresponding cooling process much like the magneto caloric effect based on the modification of  interaction parameter based on the analogy between Eqs. (\ref{eq:MCE}) and (\ref{eq:CMCE}). However, before we discuss the interaction induced cooling, let us first review the standard magnetic refrigeration.


In the magnetic refrigeration cycle, depicted in Fig.~\ref{fig:schematic_MCE}, there are four processes. $A \to B$: initially randomly oriented magnetic moments are aligned by a magnetic field, resulting in the heating of the magnetic material; $B \to C$: heat is removed from the medium to the hot bath by coupling the working medium and the hot bath; $C \to D$: by removing away the magnetic field adiabatically, the magnetic moments are randomized, that leads to cool the material below the cold ambient temperature, and finally $D \to A$: heat is extracted from the cold ambient  to the working medium by coupling the working medium and the target. This technique is also referred to  the adiabatic demagnetization refrigerator (ADR) as being shown in  the key step  $C \to D$, where the working system is demagnetized adiabatically. In the process $B \to C$ the total heat absorbed by the ambient from the working system is the area of the curved trapezoid $BCEF$, i.e. $\Delta Q_1=S_{BCEF}$, and similarly in the process $D \to A$ the total heat that is  absorbed  from the target system is $\Delta Q_2=S_{ADEF}$. The cooling efficiency of the refrigerator is $\eta=\Delta Q_2/\Delta Q_1$, which imply that the limitation of the efficiency is $\eta_{\text{max}}=T_{\text{tar}}/T_{\text{sour}}$ when $T_B \to T_c$ and $T_A \to T_D$. But the maximum of the efficiency means the minimal of the power. For a realistic application, one has to weight  the efficiency  and the power. In low temperature physics, the reachable low temperature limit is the most important issue for  engineering refrigeration. We shall further discuss the possible lowest temperature for a  cooling by near a  quantum phase transition. Below, we construct the analogous magnetic refrigeration cycle based on changing interaction parameters. 


To make concrete the above statement, we demonstrate quantum refrigeration based on the Bethe ansatz solution of the Lieb-Liniger model,  which describes the 1D Bose gas with a contact interaction. 
The Hamiltonian 
of the Lieb-Liniger model  in a 1D box with length $L$ is given by \cite{Lieb_PR_1963}
\begin{align}
\hat{H}=-\frac{\hbar^{2}}{2m}\sum_{i=1}^{N}\frac{\partial^{2}}{\partial
x_{i}^{2}}+2c\sum_{1\leq i<j\leq
N}\delta(x_{i}-x_{j}).
\label{eq:Hailtonian_first_quantization}
\end{align} 
where $m$ is the mass of the particles, $c$ is the coupling  strength which
is determined by the 1D scattering length $c =-2\hbar^2/ma_{1D}$. 
The scattering length is given by $a_{1D}=(-a_\perp^2/2a_s)[1-C(a_s/a_\perp)]$
\cite{Olshanii_PRL_1998,Dunjko_PRL_2001,Olshanii_PRL_2003}.

Before analysing  the refrigerator cycle, we first briefly study  the thermodynamic scaling invariance of this model. In order to prove  the scaling invariant nature  in  the entropy,  here we extend our discussion presented in the last section to  the exactly solved model.  The Hamiltonian (\ref{eq:Hailtonian_first_quantization}) can be solved by Bethe ansatz \cite{Lieb_PR_1963,Lieb_PRL_1968}, we list some related  key results in \cite{Sup}.   Suppose that  we have obtained the solution of the thermodynamic Bethe ansatz equation (\ref{TBA}) of dressed energy $\epsilon(k)$ under the input parameters $\mu$, $T$ and $c$, it is obvious that the dressed energy 
$\epsilon^{\prime} (k^{\prime}) = \me^{-2 \lambda} \epsilon(\me^{\lambda} k^{\prime})$ is the corresponding scaling form  for input parameters under such rescaling   $T^{\prime} = \me^{-2\lambda}T, \mu^{\prime} = \me^{-2\lambda} \mu$ and $c^{\prime} = \me^{-\lambda} c $. 
Strictly speaking, the dressed energy is a homogeneous function with 
$\epsilon(\me^{\lambda}k,\me^{-2\lambda}\mu,\me^{-2\lambda}T, \me^{-\lambda}c) = \me^{-2\lambda}\epsilon(k,\mu,T,c)$ for $\forall \lambda \in R$. By definition \cite{Sup},  the pressure can be  obtained in a straightforward way, i.e.  we may obtain an homogeneous form of the pressure  $p(\me^{-2\lambda}\mu,\me^{-2\lambda}T, \me^{-\lambda}c) = \me^{-3\lambda}p(\mu,T,c)$. By differentiation,  the density is given by  $n(\me^{-2\lambda}\mu,\me^{-2\lambda}T, \me^{-\lambda}c) = \me^{-\lambda}n(\mu,T,c)$. 
 Furthermore, the  the entropy  density $s = S/L$ is given by 
\begin{align}
s(\me^{-2\lambda}\mu,\me^{-2\lambda}T, \me^{-\lambda}c) = \me^{-\lambda}s(\mu,T,c).
\label{eq:entropy_scaling}
\end{align} 
For the system with the fixed particle number, we need $L \to L^{\prime} = \me^{\lambda} L$ to ensure 
$N^{\prime} = N$ under scaling transformation
$\mu^{\prime} = \me^{-2\lambda} \mu,\, T^{\prime} = \me^{-2\lambda} T,\,  c^{\prime} = \me^{-\lambda} c$, then according to (\ref{eq:entropy_scaling}) we arrive at the conclusion that under this scaling transformation the entropy is unchanged $S^{\prime} = S$ which  is the key conclusion we have claimed  to obtain the identity of \gps (\ref{eq:identity}) in he last section.

Similar discussions can be given on Gaudin-Yang model (see next section) and other integrable models. However, we emphasize that the identity (\ref{eq:identity}) does not depend on any particular model and has its origin in the scaling properties of the spectrum.  Historically, the study of the \gp started, in fact, from the discussion on homogeneity of thermodynamic quantities as functions of the oscillation frequency $\omega_0$ in the simple Einstein model \cite{Gruneisen_AdP_1908,Gruneisen_AdP_1912} (for details, see supporting material \cite{Sup}). 

\begin{figure*}
\includegraphics[width=0.90\textwidth]{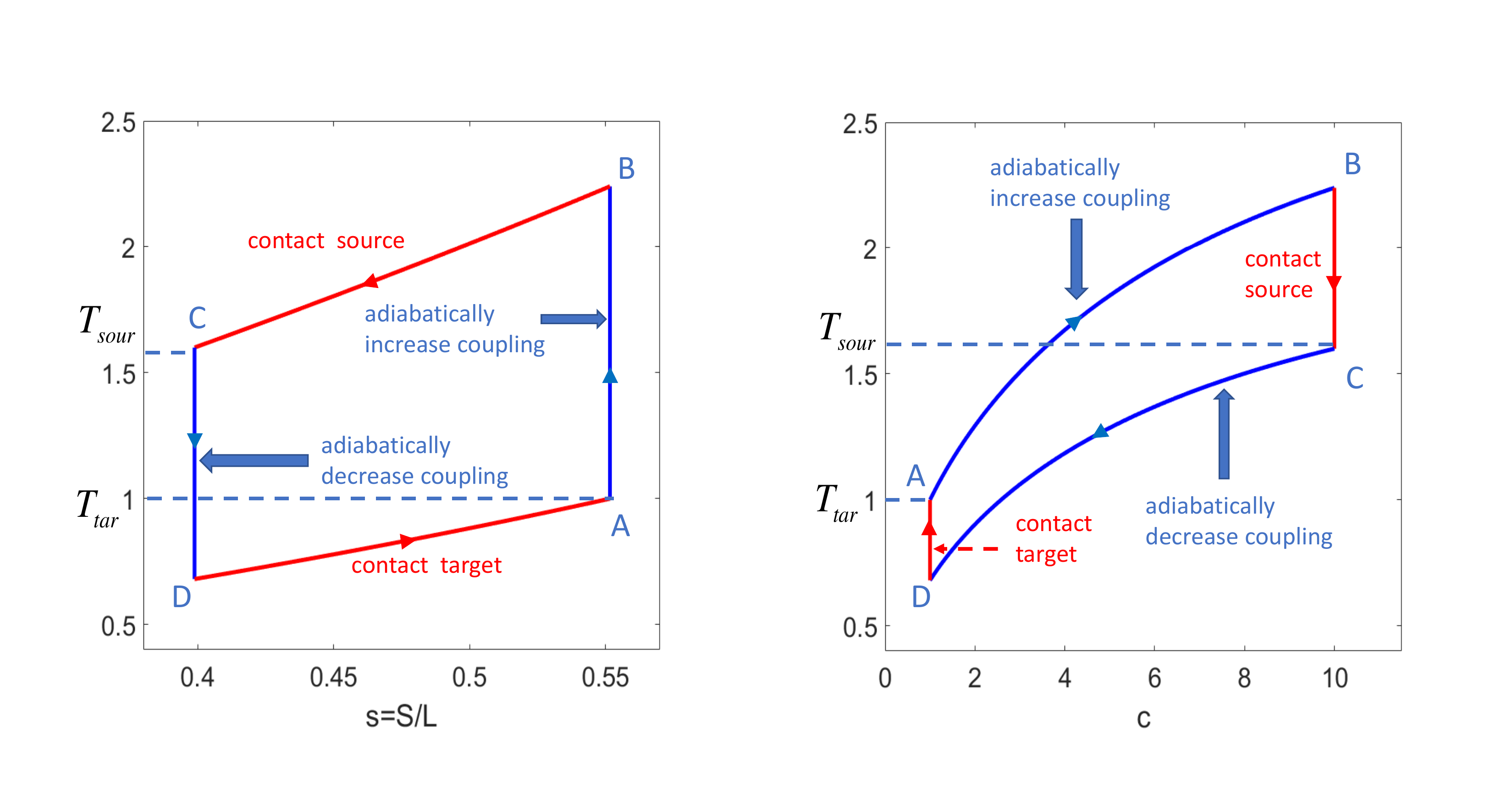}
\caption{ 
Schematic demonstration of   the interaction driven  refrigeration in the $T-s$ and $T-c$ planes. 
The cycle with four processes  is an  analog to the magnetic refrigeration which 
we discussed in the Fig.~\ref{fig:schematic_MCE}. Here the processes:   $A\to   B$: interaction ramp-up isentrope; 
 $ B \to C$: an isochore by contacting a hot source  (release  heat to the  hot source); 
$C \to  D$:  interaction ramp-down isentrope; 
$ D \to A$: a isochore by contacting a cold source (absorb  heat from  the target source). 
The cycle  is plotted by numerically 
solving the thermodynamic Bethe ansatz equations (TBAE) of the Lieb-Linger model, see  \cite{Sup}.  The cycle begins
at point A with  $n = 0.1, c = 1.0, T = 1.0$, (here all the quantities are in the nature   units  $\hbar = 2m = k_B = 1$.) then the coupling strength is tuned to strong interacting region $c = 10.0$, after contacting with the heat source the coupling strength is tuned back to $c = 1.0$, finally the working material contacts sufficiently to the target, and the cycle is finished.  
 }
\label{fig:CMCE}
\end{figure*}

Now let us return to our discuss on refrigerator cycle driven by the interaction strength $c$ in the Lieb-Liniger model (\ref{eq:Hailtonian_first_quantization}). As a direct analogy to Fig.~\ref{fig:schematic_MCE}, the interaction driven refrigerator cycle is showed in the $T-S$ and $T-c$ plane in Fig.~\ref{fig:CMCE} via rigorous calculation by thermodynamic Bethe ansatz equations, see \cite{Sup}. The implementation of such as cycle by dunning the interaction in the 1D interacting Bose gas much likes the interaction driven heat engine proposed in \cite{Chen_2018}.

The  right panel of Fig.~\ref{fig:CMCE} shows the four strokes in a cooling cycle with the interacting bosons.    $A\to   B$: The working medium  is initially in the thermal state $A$ determined by the interaction strength $c_A=1$ and temperature $T_{\rm tar}=1$.  The isentropic ramp-up of interaction takes place and the interaction strength is finally enhanced to the value $c_B$. After a quasi-adiabatic unitary evolution, the system reaches to a state with temperature $T_B$. $B \to C$: Keeping $c_B$ constant, the working medium  is coupled to the hot ambient  at  temperature $T_{\rm sour}$  and reaches the equilibrium state $(c_B,T_{\rm sour})$. The heat $\Delta Q_1$  is removed from the working medium into the hot ambient. $C \to  D$:  The working medium is decoupled from the hot ambient. By performing a  work, interaction ramp-down isentropic process takes place. 
The interaction strength decreases from $c_B$ to $c_D=c_A$, the working medium reaches the temperature $T_D$. 
$D \to A$: The working medium is coupled to the target  cold reservoir  keeping the interaction strength constant until it reaches the thermal state $(c_A,T_{\rm tar})$. The heat $\Delta Q_2$ is extracted from the target reservoir. The cooling efficiency is $\eta =\Delta Q_2/\Delta Q_1$. 

We would like to stress that in a realistic cycle, $A\to B$ and $C\to D$ are  likely not to  be rigorously adiabatic. However, the temperatures at the non-thermal state $B$ and $D$ can be still well defined if the spectra of the working system are scaling invariant, see discussion in Ref.\cite{Chen_2018}. 
The exact solution of the working system allows us to determine working efficiency in this particular case. 
We would like to mention that the modulation of the coupling strength in an interaction-driven cooling cycle can be associated with the coupling to external degrees of freedom, also see a recent study of the quantized refrigerator \cite{Niedenzu_2019}.
 In next section, we shall discuss  the reachable lowest temperature for  an engineering refrigeration with the 1D interacting fermions.

\begin{figure}[ht]
\begin{center} 
\includegraphics[width=0.5\textwidth]{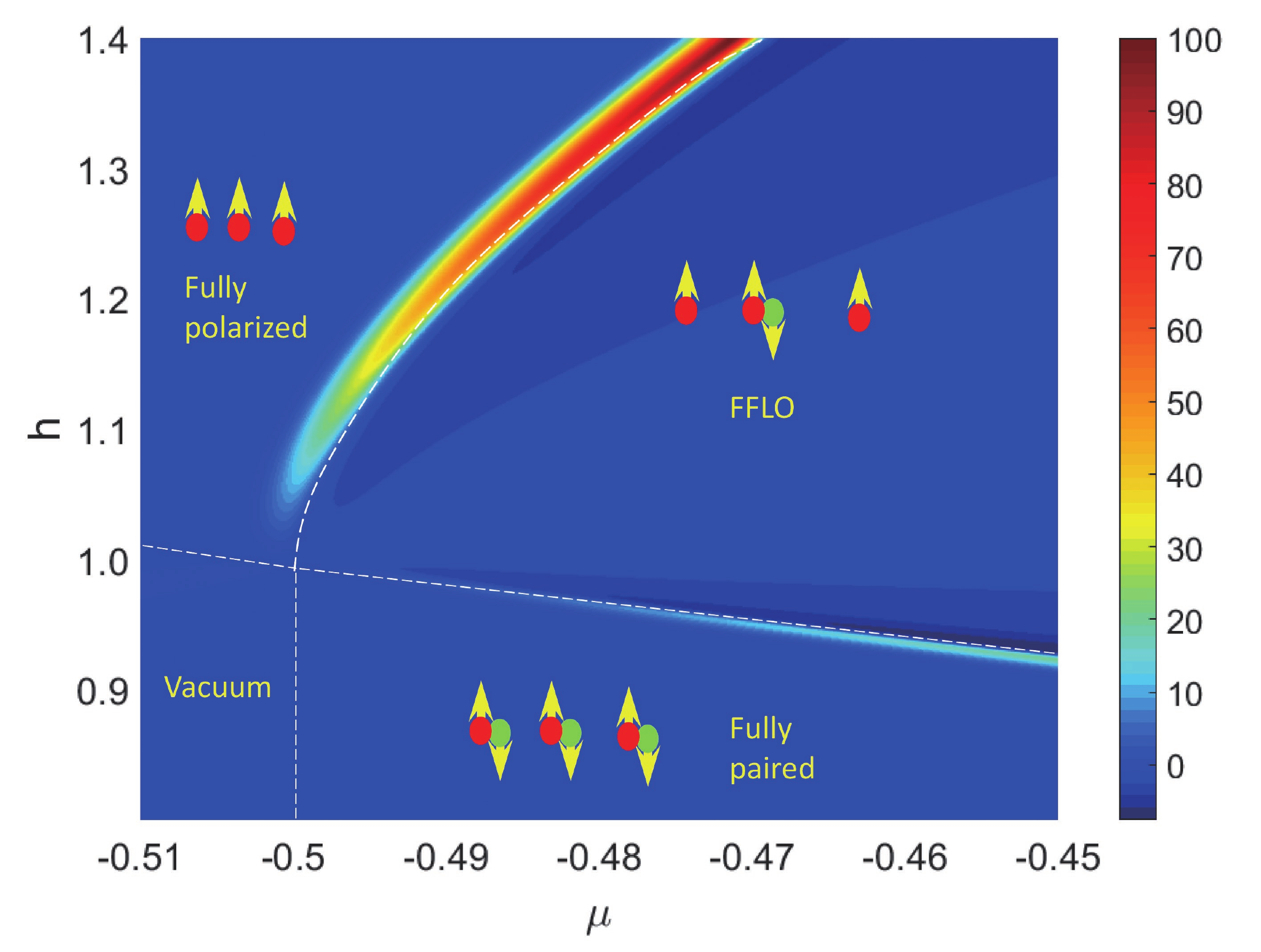}
\end{center}
\caption{Contour plot of the negative GP (\ref{eq:calculation_for_Gamma}), i.e. $-\Gamma$,  mapping out the full phase diagram of the Yang-Gaudin model with an attractive interaction in $h-\mu$ plane. It consists of three novel phases, fully paired state, Fully polarized state and a FFLO like state. 
Here  the dimensionless temperature  $t=0.001$. The GP has a sudden enhancement near the phase boundaries, giving a universal divergent scaling $\Gamma
\sim  t^{-1/2} $, see the main text. }
\label{fig:cross_3}
\end{figure}

{\bf 2. The \gp at quantum criticality. }
As discussed in the last section, the \gps play the central role in this cooling process based on the equations (\ref{eq:MCE}) or (\ref{eq:CMCE}). Since the \gps are second order derivatives with respect to free energy, it is expected that the \gps will show divergent and scaling behaviors at the QCPs~\cite{Weickert_PRB_2012,Steppke_Science_2013,
Wolf_IJMPB_2014,Ryll_PRB_2014,StraBel_PRB_2015,Gegenwart_RPP_2016,Mousumi_JPD_2016}, leading to much enhanced effects for quantum refrigeration.


In order to  illustrate this idea and to analyse  the scaling behaviors of the GPs, we take the Yang-Gaudin  model  \cite{Yang_PRL_1967,Gaudin_1967} as an  example to carry out  rigorous calculations. This model  is  one of the most important exactly solvable quantum many-body systems. It was solved long ago by Yang \cite{Yang_PRL_1967} and Gaudin \cite{Gaudin_1967} using the Bethe ansatz. 
Theoretical predictions for the existence of a Fulde-Ferrell-Larkin-Ovchinnikov (FFLO) pairing state in the 1D interacting Fermi gas have emerged by using the exact solution \cite{Orso:2007,Hu:2007,Guan:2007}. 
The key features of this $T=0$ phase diagram were experimentally confirmed using finite temperature density profiles of trapped fermionic ${}^6$Li atoms \cite{Liao_Nature_2010}.
The Hamiltonian of the Yang-Gaudin model 
\begin{eqnarray}
{\hat{H}} &=& \sum _{\sigma=\downarrow,\uparrow} \int \phi _{\sigma}^{\dagger}(x) \left
(-\frac{\hbar^{2}}{2m}\frac{d^{2}}{dx^{2}} + \mu_{\sigma} \right ) \phi
_{\sigma}^{}(x) \mr x 
\notag \\
& &+  g_{\rm 1D} \int \phi _{\downarrow}^{\dagger}(x) \phi
_{\uparrow}^{\dagger}(x) \phi _{\uparrow}^{}(x)
\phi _{\downarrow}^{}(x) \mr x\nonumber\\
&&
 - \frac12{h} \int \left (\phi _{\uparrow}^{\dagger}(x) \phi
_{\uparrow}^{}(x) - \phi _{\downarrow}^{\dagger}(x) \phi
_{\downarrow}^{}(x) \right ) \mr x 
\label{eq:Hailtonian_Gaudin_Yang}
\end{eqnarray}
describes a 1D $\delta$-function interacting two-component 
Fermi gas of $N$ fermions with mass $m$ and an external magnetic field $H$
constrained by periodic boundary conditions to a line of length $L$. 
Where   $g_{1D}=-2\hbar^2/(ma_{1D})$ is determined by an effective scattering length $a_{1D}$ via Feshbach
resonances or confinement-induced resonances~\cite{Olshanii_PRL_1998,Dunjko_PRL_2001,Olshanii_PRL_2003}. $g_{1D}>0$ ($<0$) represents repulsive (attractive) interaction.
Usually $c=mg_{1D}/\hbar ^{2}=-2/a_{1D}$  denotes  the effective interaction strength.

Here we show that the different GPs  (\ref{eq:calculation_for_Gamma}), (\ref{eq:calculate_mag_gamma}) and (\ref{eq:definition_coupling_gruneisen}) not only signal quantum phase transitions but also quantify various fluctuations  in quantum systems. 
Using the exact TBA equations, a  full critical phase digram of the Yang-Gaudin model at $t=0.0001\epsilon_b$ is determined by the  the GP expression  (\ref{eq:calculation_for_Gamma}), see supplementary material \cite{Sup}. 
In this contour plot the rescaled units were used, i.e. $\tilde{t}=t/(c^2/2)$, $ \tilde{\mu}=\mu/(c^2/2)$ and $\tilde{h}=h/(c^2/2)$. 
We observe that the GP  (\ref{eq:calculation_for_Gamma}) characterizes the universal  divergent scaling near the phases boundaries. 
It shows  that the  energy-pressure covariance has a stronger fluctuations than the energy fluctuation. 
This nature can be used to identify different quantum phases, i.e. novel Luttinger liquids  of  fully-paired state,  FFLO-like pairing state  and fully polarized state  are determined at low temperatures, see Fig.~\ref{fig:cross_3}.
We show that the phase boundaries  between the   fully polarized phase and FFLO-like pairing phase and between  the fully paired phase and FFLO-like pairing phase in Fig.~\ref{fig:cross_3} can be cast into a  universal scaling (for a constant $h$)
\begin{eqnarray}
\Gamma
=\lambda  t^{-1/2} 
\mathcal{G}\left(\frac{2(\mu-\mu_{c})}{t}\right)
\label{eq:scalings}
\end{eqnarray}
with the factor $\lambda =\sqrt{\pi}n$ and $\lambda =\sqrt{2\pi} n$, respectively. 
In the above equation, $n$ is the density, $\mathcal{G}(x)$ is the scaling function and $\mu$ is an  effective chemical potential \cite{Guan_RMP_2013}. 
More detailed study on the quantum scalings of the GPs  (\ref{eq:scalings}) will be published elsewhere.
In addition, the magnetic and interacting GPs (\ref{eq:calculate_mag_gamma}) and (\ref{eq:definition_coupling_gruneisen})  also give the  full phase diagram at low temperatures. 

The divergence of the GPs at $T \to 0$ near QCPs can be clearly understood by investigating the entropy of the system.  At low temperatures, the state of the  system away from the critical points  usually behaves like  Fermi liquid (or Tomonaga-Luttinger liquid region in 1D), see Fig.~\ref{fig:cross_3}.  The entropy $S \propto T$.  In contrast, the entropy at the quantum criticality behaves as \cite{Guan_PRA_2011,Fisher_PRB_1989}
\begin{align}
\frac{S}{V} \propto T^{(d/z)+1-(1/\nu z)} \mathcal{K}(\frac{\mu - \mu_c}{T^{1/\nu z}}).
\label{eq:scaling_function}
\end{align}
For 1D system, $z=2$ is  the dynamic critical exponent and $\nu=1/2$ is the critical index for correlation length,  
whereas the $\mu$ presents an  effective chemical potential, and the $\mu_c$ is the quantum critical point. 
$\mathcal{K}(x)$ is some analytica scaling function. 
Note that the entropy is exactly zero at zero temperature, so there is no background term in \cref{eq:scaling_function}  \cite{Guan_RMP_2013,Jiang_CPB_2015,Yang_NC_2014,He_PRB_2017,Yu_PRB_2016,Cheng_PRB_2018}.
For the  Gaudin-Yang model, the entropy $S \propto \sqrt{T} \gg T$ near the QCPs, which implies the local maximum of the entropy at QCPs.  If we plot the entropy in the $T-H$ plane, 
see Fig.~\ref{fig:contourplot_entropy}, the isentropic lines will be bent down significantly at QCPs. According to equation (\ref{eq:MCE}), the GPs are proportional to the slope of the isentropic line which leads to the divergence of the GP when $T \to 0$.

The local maximum of the entropy at low temperature reveals the essence of the QCPs when the low lying excitation  becomes degenerate with the ground state \cite{Sachdev_2001}.  In general, the divergence of the GPs is also present in generic models when quantum phase transition occurs. This nature has been extensively studied both in theory and experiments \cite{Zhu_PRL_2003,Kuchler_PRL_2003,Garst_PRB_2005,Kuchler_PRL_2006,Kuchler_STAM_2007,Tokiwa_PRL_2009, Wolf_PNAS_2011,Smith_AEM_2012,Weickert_PRB_2012,Steppke_Science_2013,Ryll_PRB_2014,StraBel_PRB_2015, Gegenwart_RPP_2016,Lee_PRB_2017,Watanabe_JPSJ_2018}.   

%

{\bf 3. Quantum refrigeration near a phase transition.} For refrigeration, it is important to ask what the lowest temperature one can achieve. In Supplementary material \cite{Sup}, we answer the above question for the free Fermi gas. This question seems trivial in the common refrigeration \cite{Smith_AEM_2012}. However,  if the system approaches its quantum critical point, things can be significantly different. The divergent behaviour of the GPs near QCPs can lead to significant cooling of the system. In fact, the feature of local maximum of the entropy leads  to a  local temperature minimum  in an  isentropic process, see Fig.\ref{fig:contourplot_entropy}. Consequently, one can  take this  advantage of the quantum phase transition  to enhance the MCE (or interaction driven MCE).
Using exact solution of the 1D attractive Fermi gas, we further demonstrate  a magnetic (or interaction driven)  refrigeration in the    interacting Fermi gas \cite{Yang_PRL_1967,Gaudin_1967} .

%
%

\begin{figure}[h]
\begin{center}
\includegraphics[width=0.5\textwidth]{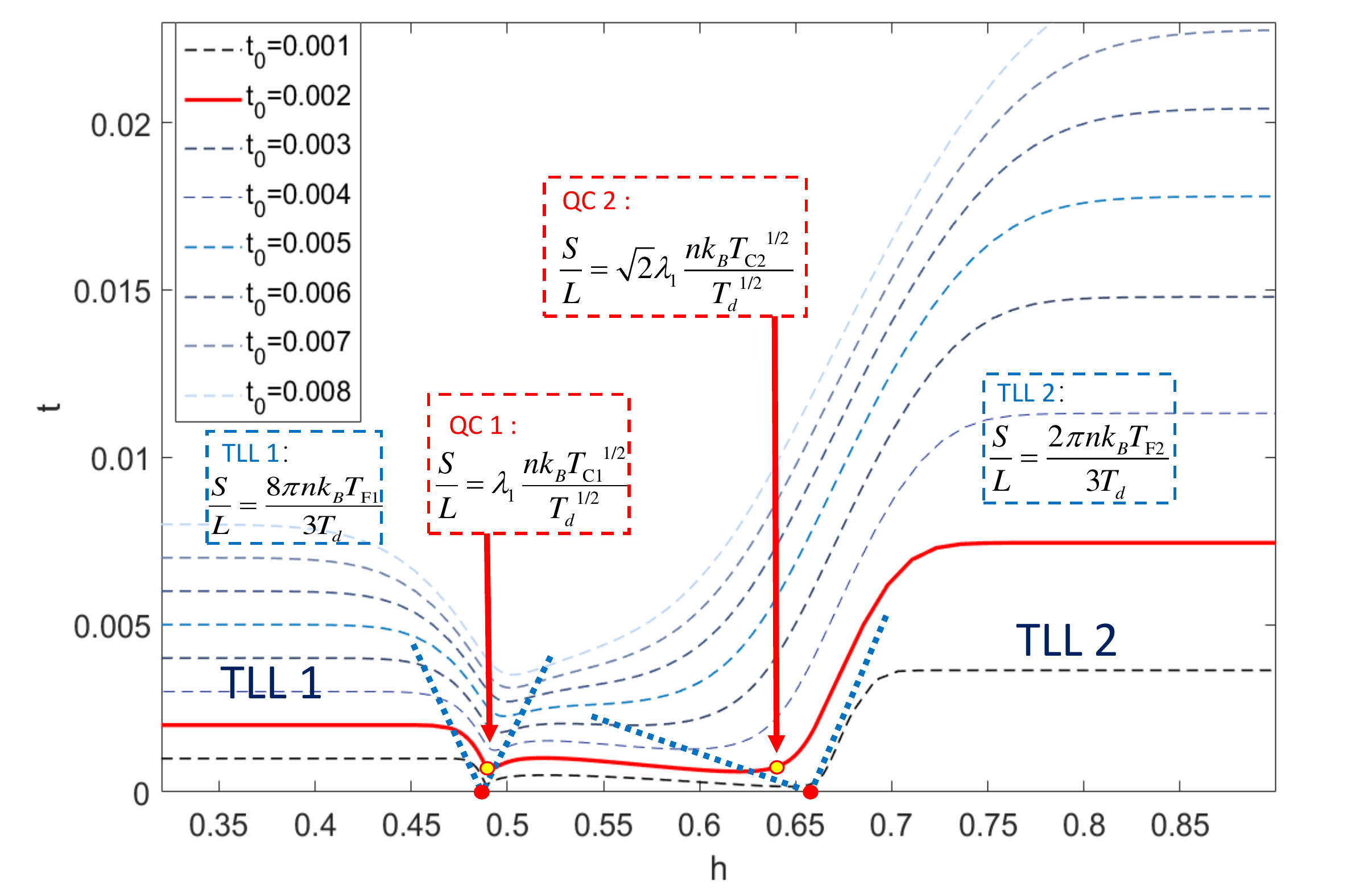}
\end{center}
\caption{The contour plot of the entropy in $t-h$ plane  for the attractive Yang-Gaudin  model at low temperatures.   Here the magnetic field 
$h = H/\epsilon_b$ and the  temperature $t = T/\epsilon_b$  are rescaled by the binding energy with $2m = \hbar = k_B = 1$.  We carried out  our calculation through the TBA equations \cite{Sup} with a fixed density $n = 0.1$.  $h_{c1} $ and $h_{c2} $ are the critical points for the phase transitions from fully paired TLL to the FLLO like phase and from the FFLO like phase to the fully polarized phase at $t=0$,  respectively. 
The dash lines in  different colors present the contour values of entropies at different temperatures. 
The bending down of the contour lines  indicates  an entropy accumulation with a minimum temperature ( yellow dot).  
For $h<h_{c1}$ the system is in the  TLL of bound pairs obeying  the state equation (\ref{entropy_two_TLL_phase}), whereas for $h>h_{c2}$ the system is in a  fully polarized TLL obeying the equation  (\ref{entropy_two_TLL_phase}).  These  analytical results of the state equations directly give  the minima of the temperature  during the adiabatic demagnetization processes,  see  (\ref{eq:cool_down_equation-YG}).
}
\label{fig:contourplot_entropy}
\end{figure}

%
%
 %

Using  the condition  (see the equation (\ref{eq:MCE}))
\begin{equation}
\Gamma_{\text{mag}}=0,
\end{equation}
we may answer the above question .
For  the  Yang-Gaudin  model \cite{Sup},  we expect an enhancement of the cooling  efficiency  when the working system is approaching to a quantum critical point in the phase diagram Fig.~\ref{fig:cross_3}.
Here we focus on the low  temperature region, i.e. $T\ll T_d$, here $k_B T_d=(\frac{\hbar^2n^2}{2m})$ is the degenerate temperature. 
Fig.~ \ref{fig:contourplot_entropy} shows that the condition  $\Gamma_{\text{mag}}=0$  gives solutions for each  quantum phase transition. Like the free Fermi gas given in   \cite{Sup}, the condition $\Gamma_{\text{mag}}=0$ leads to two independent equations for  the Yang-Gaudin model at the two quantum critical points, namely, 
\begin{align}
-\frac{1}{2}\Li_{\frac{1}{2}}(-\me^{\tA^{(r)}/t})+\frac{\tA^{(r)}}{t}
\Li_{-\frac{1}{2}}(-\me^{\tA^{(r)}/t})=0, 
\label{eq:gaudinyang_zero_mag_gp}
\end{align}
where $r=1$ and $r=2$ stand for the unpaired fermions and bound pairs, respectively. 
This means that at the phase transition $H_{c1}$ the density of state of unpaired fermions is suddenly changed, whereas at the critical point $H_{c2}$ the density of state of the paired fermions is suddenly changed. 
While the effective chemical potentials  of unpaired fermion and pairs,  here $\tA^{(1)} = (\mu+H/2)/\epsilon_b, \tA^{(2)} = (2\mu+c^2/2)/\epsilon_b $ were rescaled by the bonding energy $\epsilon_b=c^2/2$.
Here  $\mu$ is the chemical potential, $t = T/\epsilon_b$ is the rescaled  temperature. 

Equation (\ref{eq:gaudinyang_zero_mag_gp}) is very similar to the equation $\mathcal{Y}(x)=x-\frac{\Li_{1/2}(-\me^{x})}{2\Li_{-1/2}(-\me^{x})}=0$ found in  the free Fermi gas~\cite{Sup}.
We thus have the same solution $\tA^{(r)}/t=x_0 \approx 1.3117$. 
Substituting  this solution into TBA results given in \cite{Sup}, we get entropies at the phase transitions point from a fully-paired phase to the FFLO-like state and from the FFLO liked state to the fully paired Fermi gas, respectively 
\begin{eqnarray}
\frac{S}{L} &=& \lambda_1 \cdot \frac{\sqrt{m}}{\hbar\sqrt{2\pi}} k_B^{3/2} T_{c1}^{1/2}, \quad  \text{for } \,\,H \to  H_{m1}\nonumber 
 \\
\frac{S}{L}& =& \lambda_1 \cdot \frac{\sqrt{m}}{\hbar\sqrt{\pi}} k_B^{3/2} T_{c2}^{1/2}, \quad  \text{for  } \,\,H \to  H_{m2},
\label{entropy_two_critical_point}
\end{eqnarray}
where $\lambda_1=x_0\Li_{1/2}(-\me^{x_0})-\frac{3}{2}\Li_{3/2}(-\me^{x_0})\approx 1.3467$.
 $H_{m1}$ and $H_{m2}$ are two critical fields corresponding to the two  temperature minima in the isentropic contour  lines. 
Using  TBA equation,  we have the entropy in the liquid phases of pairs and fully-polarized fermions
\begin{eqnarray}
\frac{S}{L}& =& \frac{4m}{3\hbar^2} k_B^2 T_{L1} n^{-1},\quad 
 \text{for  }\,\,  H < H_{m1}, \nonumber
 \\
\frac{S}{L} &=& \frac{m}{3\hbar^2} k_B^2 T_{L2} n^{-1}, \quad 
\text{for  }\,\,  H > H_{m2}.
\label{entropy_two_TLL_phase}
\end{eqnarray}
Here $T_{L1}$ and  $T_{L2}$ are the temperatures  in the Luttinger liquid regions, see the phases $TLL1$ and $TLL2$ in Fig.~\ref{fig:contourplot_entropy}.
For  the first equation in  (\ref{entropy_two_TLL_phase}),  we  applied the strong coupling  condition 
$\gamma = c/n \gg 1$. 
From these equations (\ref{entropy_two_critical_point}) and (\ref{entropy_two_TLL_phase}),  we find  two temperature minima of the refrigeration around the two phase transitions 
\begin{eqnarray}
\frac{T_{c1}}{T_{d}}&=& 8\lambda_2^2 \cdot\left( \frac{T_{L1}}{T_d}\right)^2,\nonumber
\\
\frac{T_{c2}}{T_{d}} &=& \frac{\lambda_2^2}{2} \cdot \left( \frac{T_{L2}}{T_d}\right)^2
\label{eq:cool_down_equation-YG}
\end{eqnarray} 
with  $\lambda_2 = \frac{2\pi}{3\lambda_1} \approx 1.5552$.
We further observe that   the leading  contribution to the entropy at the  critical point $H_{m1}$  involves  the excitations   of the excess fermions  \cite{Yu_PRB_2016}. 
Whereas at  the critical point $H_{m2}$, the leading  contribution to the entropy comes from  the excitations of the bound  pairs. 
  In the  isentropic process, the system thus  can retain more entropy per unit temperature near  the finite temperature critical point $H_{m2}$.
This result reveals  an enhancement of   the cooling  efficiency.   In cold atom experiment, the temperature  is usually much lower than the degenerate temperature \cite{Pethick_Cambridge_2008}, i.e.  $T_{L1}/T_d \ll 1$ and $T_{L2}/T_d \ll 1$. Trom equation (\ref{eq:cool_down_equation-YG}), we thus have $T_{c1} \ll T_{L1}$ and $T_{c2} \ll T_{L2}$.  Moreover,  the ideal limit of the temperature $T_{c_{1,2}}$ for the refrigeration is twice in order of magnitude compared to the temperature of the heat source $T_{L_{1,2}}$.

\section{IV. Summary}
We have conducted a comprehensive investigation of the GP for ultracold quantum gases, including its origin, new identity, caloric effects and quantum refrigeration. We have proposed the interaction related GP which reveals characteristic  energy scales  of quantum system induced by the variation of the interaction. Together with the other two GPs related to the variations of volume and magnetic field, we have established a new identity between them which characterises the universal scalings of fluctuations and caloric effect in quantum gases. Based on the entropy accumulation at the quantum critical point,  two promising protocols of quantum refrigeration driven either by interaction or by magnetic field have been studied. Using Bethe Ansatz, we studied  the expansionary, the  magnetic and the interacting GPs,  quantum refrigeration, magnetocaloric effect and quantum critical phenomenon of the Lieb-Liniger model and Yang-Gaudin model. Our method opens to further  study the GPs and quantum refrigeration for   quantum gases of ultracold atoms with different spin symmetries  in 1D and higher dimensions.

%
%


\section*{Acknowledgement}
The author thanks Y.X. Liu, L. Peng, Y.Z. Jiang, Y.Y. Chen, F. He, H. Pu and R. Hulet  for helpful discussions. 
This work is supported by   the National Key R\&D Program of China  No. 2017YFA0304500,  the key NSFC grant No.11534014 and No. 11804377.
YCY thank the University of Hong Kong for a kind hospitality. XWG acknowledge Rice University for a support of his visit.

\bibliographystyle{unsrt}

\bibliography{Gruneisen_bib}

\begin{thebibliography}{10}

\bibitem{Gruneisen_AdP_1908}
E.~Gr\"uneisen.
\newblock {{\"Uber} die Thermische Ausdehnung und die Spezifische Wärme der
  Metalle}.
\newblock {\em Annalen der Physik}, 331(6):211--216, 1908.

\bibitem{Gruneisen_AdP_1912}
E.~Gr\"uneisen.
\newblock {Theorie des Festen Zustandes Einatomiger Elemente}.
\newblock {\em Annalen der Physik}, 344(12):257--306, 1912.

\bibitem{Stacey_PEPI_1995}
F.~D. Stacey.
\newblock Theory of thermal and elastic properties of the lower mantle and
  core.
\newblock {\em Physics of the Earth and Planetary Interiors}, 89(3):219--245,
  1995.

\bibitem{Shanker_PEPI_2017}
J.~Shanker, K.~Sunil, and B.~S. Sharma.
\newblock The {Gr\"uneisen} parameter and its higher order derivatives for the
  earth lower mantle and core.
\newblock {\em Physics of the Earth and Planetary Interiors}, 262:41--47, 2017.

\bibitem{Mausbach_JCP_2016}
P.~Mausbach, A.~Koster, G.~Rutkai, M.~Thol, and J.~Vrabec.
\newblock Comparative study of the {Gr\"uneisen} parameter for 28 pure fluids.
\newblock {\em Journal of Chemical Physics}, 144(24), 2016.

\bibitem{Liu_PCCP_2017}
G.~Liu, J.~Zhou, and H.~Wang.
\newblock Anisotropic thermal expansion of {SnSe} from first-principles
  calculations based on {Gr\"uneisen's} theory.
\newblock {\em Physical Chemistry Chemical Physics}, 19(23):15187--15193, 2017.

\bibitem{Wang_PRE_2017}
L.~Wang, M.~T. Dove, K.~Trachenko, Y.~D. Fomin, and V.~V. Brazhkin.
\newblock Supercritical {Gr\"uneisen} parameter and its universality at the
  {Frenkel} line.
\newblock {\em Physical Review E}, 96(1), 2017.

\bibitem{Kumar_PJP_2016}
S.~Kumar, S.~K. Sharma, and O.~P. Pandey.
\newblock Brief report: volume dependence of {Gr\"uneisen} parameter for solids
  under extreme compression.
\newblock {\em Pramana-Journal of Physics}, 87(2), 2016.

\bibitem{Khrapak_PP_2017}
S.~A. Khrapak.
\newblock Gr\"uneisen parameter for strongly coupled {Yukawa} systems.
\newblock {\em Physics of Plasmas}, 24(4), 2017.

\bibitem{Kuchler_PRL_2003}
R.~Kuchler, N.~Oeschler, P.~Gegenwart, T.~Cichorek, K.~Neumaier, O.~Tegus,
  C.~Geibel, J.~A. Mydosh, F.~Steglich, L.~Zhu, and Q.~Si.
\newblock Divergence of the {Gr\"uneisen} ratio at quantum critical points in
  heavy fermion metals.
\newblock {\em Physical Review Letters}, 91(6), 2003.

\bibitem{Kuchler_PRL_2006}
R.~K\"uchler, P.~Gegenwart, J.~Custers, O.~Stockert, N.~Caroca-Canales,
  C.~Geibel, J.~G. Sereni, and F.~Steglich.
\newblock Quantum criticality in the cubic heavy-{Fermion} system
  {${\mathrm{CeIn}}_{3\ensuremath{-}x}{\mathrm{Sn}}_{x}$}.
\newblock {\em Physical Review Letters}, 96(25):256403, 2006.

\bibitem{Kuchler_STAM_2007}
R.~K\"uchler, P.~Gegenwart, C.~Geibel, and F.~Steglich.
\newblock Systematic study of the {Gr\"uneisen} ratio near quantum critical
  points.
\newblock {\em Science and Technology of Advanced Materials}, 8(5):428--433,
  2007.

\bibitem{Gegenwart_RPP_2016}
P.~Gegenwart.
\newblock Gr\"uneisen parameter studies on heavy fermion quantum criticality.
\newblock {\em Reports on Progress in Physics}, 79(11), 2016.

\bibitem{Steppke_Science_2013}
A.~Steppke, R.~K\"uchler, S.~Lausberg, E.~Lengyel, L.~Steinke, R.~Borth,
  T.~L\"uhmann, C.~Krellner, M.~Nicklas, C.~Geibel, F.~Steglich, and M.~Brando.
\newblock Ferromagnetic quantum critical point in the heavy-{Fermion} metal
  {\({y}bni_4(p_{1-x}as_x)\)}.
\newblock {\em Science}, 339(6122):933--936, 2013.

\bibitem{Tokiwa_PRL_2009}
Y.~Tokiwa, T.~Radu, C.~Geibel, F.~Steglich, and P.~Gegenwart.
\newblock Divergence of the magnetic {Gr\"uneisen} ratio at the field-induced
  quantum critical point in ${\mathrm{ybrh}}_{2}{\mathrm{si}}_{2}$.
\newblock {\em Physical Review Letters}, 102(6):066401, 2009.

\bibitem{Wolf_PNAS_2011}
B.~Wolf, Y.~Tsui, D.~Jaiswal-Nagar, U.~Tutsch, A.~Honecker, K.~Removi{\c
  c}-Langer, G.~Hofmann, A.~Prokofiev, W.~Assmus, G.~Donath, and M.~Lang.
\newblock Magnetocaloric effect and magnetic cooling near a field-induced
  quantum-critical point.
\newblock {\em Proceedings of the National Academy of Sciences}, 108(17):6862,
  2011.

\bibitem{Wolf_IJMPB_2014}
B.~Wolf, A.~Honecker, W.~Hofstetter, U.~Tutsch, and M.~Lang.
\newblock Cooling through quantum criticality and many-body effects in
  condensed matter and cold gases.
\newblock {\em International Journal of Modern Physics B}, 28(26), 2014.

\bibitem{Ryll_PRB_2014}
H.~Ryll, K.~Kiefer, C.~Ruegg, S.~Ward, K.~W. Kramer, D.~Biner, P.~Bouillot,
  E.~Coira, T.~Giamarchi, and C.~Kollath.
\newblock Magnetic entropy landscape and {Gr\"uneisen} parameter of a quantum
  spin ladder.
\newblock {\em Physical Review B}, 89(14), 2014.

\bibitem{Chin_RMP_2010}
C.~Chin, R.~Grimm, P.~Julienne, and E.~Tiesinga.
\newblock Feshbach resonances in ultracold gases.
\newblock {\em Reviews of Modern Physics}, 82(2):1225--1286, 2010.
\newblock RMP.

\bibitem{Guan_RMP_2013}
X.-W. Guan, M.~T. Batchelor, and C.-H. Lee.
\newblock Fermi gases in one dimension: from {Bethe} ansatz to experiments.
\newblock {\em Reviews of Modern Physics}, 85(4):1633--1691, 2013.

\bibitem{Zhu_PRL_2003}
L.~J. Zhu, M.~Garst, A.~Rosch, and Q.~M. Si.
\newblock Universally diverging {Gr\"uneisen} parameter and the magnetocaloric
  effect close to quantum critical points.
\newblock {\em Physical Review Letters}, 91(6), 2003.

\bibitem{Yoshinori_JPCM_2006}
Y.~Takahashi and H.~Nakano.
\newblock Magnetovolume effect of itinerant electron ferromagnets.
\newblock {\em Journal of Physics: Condensed Matter}, 18(2):521, 2006.

\bibitem{Smith_AEM_2012}
A.~Smith, C.~R.~H. Bahl, R.~Bjørk, K.~Engelbrecht, K.~K. Nielsen, and
  N.~Pryds.
\newblock Materials challenges for high performance magnetocaloric
  refrigeration devices.
\newblock {\em Advanced Energy Materials}, 2(11):1288--1318, 2012.

\bibitem{Mousumi_JPD_2016}
S.~B. Mousumi and P.~Pankaj.
\newblock Study of magnetic entropy and heat capacity in ferrimagnetic {Fe 3 Se
  4} nanorods.
\newblock {\em Journal of Physics D: Applied Physics}, 49(19):195003, 2016.

\bibitem{Breunig_ScienceAdvance_2017}
O.~Breunig, M.~Garst, A.~Kl\"umper, J.~Rohrkamp, M.~M. Turnbull, and T.~Lorenz.
\newblock Quantum criticality in the spin-1/2 {Heisenberg} chain system copper
  pyrazine dinitrate.
\newblock {\em Science Advances}, 3(12):eaao3773, 2017.

\bibitem{Liu_SSC_2017}
Z.~J. Liu, T.~Song, X.~W. Sun, Q.~Ma, T.~Wang, and Y.~Guo.
\newblock Thermal expansion, heat capacity and {Gr\"uneisen} parameter of
  iridium phosphide {Ir2P} from quasi-harmonic {Debye} model.
\newblock {\em Solid State Communications}, 253:19--23, 2017.

\bibitem{Wilson_1975}
K.~G. Wilson.
\newblock The renormalization group: critical phenomena and the {Kondo}
  problem.
\newblock {\em Review of Modern Physics}, 47:773--840, Oct 1975.

\bibitem{Sup}
{\em See supplemental matrial}.

\bibitem{Yu_PRB_2016}
Y.-C. Yu, Y.-Y. Chen, H.-Q. Lin, R.~A. Römer, and X.-W. Guan.
\newblock Dimensionless ratios: characteristics of quantum liquids and their
  phase transitions.
\newblock {\em Physical Review B}, 94(19):195129, 2016.

\bibitem{He_PRB_2017}
F.~He, Y.-Z. Jiang, Y.-C. Yu, H.-Q. Lin, and X.-W. Guan.
\newblock Quantum criticality of spinons.
\newblock {\em Physical Review B}, 96(22):220401, 2017.

\bibitem{Landau_1980e}
L.~D. Landau and E.~M. Lifshitz.
\newblock {\em {Statistical Physics: Volume 5}}.
\newblock Butterworth-Heinemann, 1980.

\bibitem{Weickert_PRB_2012}
F.~Weickert, R.~K\"uchler, A.~Steppke, L.~Pedrero, M.~Nicklas, M.~Brando,
  F.~Steglich, M.~Jaime, V.~S. Zapf, A.~Paduan-Filho, K.~A. Al-Hassanieh, C.~D.
  Batista, and P.~Sengupta.
\newblock Low-temperature thermodynamic properties near the field-induced
  quantum critical point in {NiCl${}_{2}$-4SC(NH${}_{2}$)${}_{2}$}.
\newblock {\em Physical Review B}, 85(18):184408, 2012.

\bibitem{Garst_PRB_2005}
M.~Garst and A.~Rosch.
\newblock Sign change of the {Gr\"uneisen} parameter and magnetocaloric effect
  near quantum critical points.
\newblock {\em Physical Review B}, 72(20), 2005.

\bibitem{StraBel_PRB_2015}
D.~Stra\ss~el, P.~Kopietz, and S.~Eggert.
\newblock Magnetocaloric effects, quantum critical points, and the
  {B}erezinsky-{K}osterlitz-{T}houless transition in two-dimensional coupled
  spin-dimer systems.
\newblock {\em Physical Review B}, 91(13):134406, 2015.

\bibitem{Medley-PhysRevLett.106.195301}
P.~Medley, D.~M. Weld, H.~Miyake, D.~E. Pritchard, and W.~Ketterle.
\newblock Spin gradient demagnetization cooling of ultracold atoms.
\newblock {\em Physical Review Letters}, 106:195301, May 2011.

\bibitem{Weld-PhysRevA.82.051603}
D.~M. Weld, H.~Miyake, P.~Medley, D.~E. Pritchard, and W.~Ketterle.
\newblock Thermometry and refrigeration in a two-component {Mott} insulator of
  ultracold atoms.
\newblock {\em Physical Review A}, 82:051603, Nov 2010.

\bibitem{Yang_NC_2014}
Y.-Y. Chen, Y.-Z. Jiang, X.-W. Guan, and Q.~Zhou.
\newblock Critical behaviours of contact near phase transitions.
\newblock {\em Nature Communications}, 5:5140, 2014.

\bibitem{Olshanii_PRL_1998}
M.~Olshanii.
\newblock Atomic scattering in the presence of an external confinement and a
  gas of impenetrable {Bosons}.
\newblock {\em Physical Review Letters}, 81(5):938--941, 1998.

\bibitem{Souza_EJP_2016}
M.~de~Souza, P.~Menegasso, R.~Paupitz, A.~Seridonio, and R.~E. Lagos.
\newblock Gr\"uneisen parameter for gases and superfluid helium.
\newblock {\em European Journal of Physics}, 37(5), 2016.

\bibitem{Ekkes_JPD_2005}
B.~Ekkes.
\newblock Developments in magnetocaloric refrigeration.
\newblock {\em Journal of Physics D: Applied Physics}, 38(23):R381, 2005.

\bibitem{Khovaylo_PSS_2014}
V.~V. Khovaylo, V.~V. Rodionova, S.~N. Shevyrtalov, and V.~Novosad.
\newblock Magnetocaloric effect in “reduced” dimensions: thin films,
  ribbons, and microwires of {Heusler} alloys and related compounds.
\newblock {\em physica status solidi (b)}, 251(10):2104--2113, 2014.

\bibitem{Lieb_PR_1963}
E.~H. Lieb and W.~Liniger.
\newblock Exact analysis of an interacting {Bose} gas. {I}. {T}he general
  solution and the ground state.
\newblock {\em Physical Review}, 130(4):1605--1616, 1963.

\bibitem{Dunjko_PRL_2001}
V.~Dunjko, V.~Lorent, and M.~Olshanii.
\newblock Bosons in cigar-shaped traps: {Thomas-Fermi} regime,
  {Tonks-Girardeau} regime, and in between.
\newblock {\em Physical Review Letters}, 86(24):5413--5416, 2001.

\bibitem{Olshanii_PRL_2003}
M.~Olshanii and V.~Dunjko.
\newblock Short-distance correlation properties of the {Lieb-Liniger} system
  and momentum distributions of trapped one-dimensional atomic gases.
\newblock {\em Physical Review Letters}, 91(9):090401, 2003.

\bibitem{Lieb_PRL_1968}
E.~H. Lieb and F.~Y. Wu.
\newblock Absence of {Mott} transition in an exact solution of the short-range,
  one-band model in one dimension.
\newblock {\em Physical Review Letters}, 20(25):1445--1448, 1968.

\bibitem{Chen_2018}
Y.-Y. Chen, G.~Watanabe, Y.-C. Yu, X.-W. Guan, and A.~del Campo.
\newblock An interaction-driven many-particle quantum heat engine: universal
  behavior.

\bibitem{Niedenzu_2019}
W.~Niedenzu, I.~Mazets, G.~Kurizki, and F.~Jendrzejewski.
\newblock Quantized refrigerator for an atomic cloud.
\newblock {\em Quantum}, 3:155, jun 2019.

\bibitem{Yang_PRL_1967}
C.~N. Yang.
\newblock Some exact results for the many-body problem in one dimension with
  repulsive delta-function interaction.
\newblock {\em Physical Review Letters}, 19(23):1312--1315, 1967.

\bibitem{Gaudin_1967}
M.~Gaudin.
\newblock Un systeme a une dimension de fermions en interaction.
\newblock {\em Physics Letters A}, 24(1):55--56, jan 1967.

\bibitem{Orso:2007}
G.~Orso.
\newblock Attractive {Fermi} gases with unequal spin populations in highly
  elongated traps.
\newblock {\em Physical Review Letters}, 98:070402, Feb 2007.

\bibitem{Hu:2007}
H.~Hu, X.-J. Liu, and P.~D. Drummond.
\newblock Phase diagram of a strongly interacting polarized {Fermi} gas in one
  dimension.
\newblock {\em Physical Review Letters}, 98:070403, Feb 2007.

\bibitem{Guan:2007}
M.~T. Guan, X. W.~Batchelor, C.~Lee, and M.~Bortz.
\newblock Phase transitions and pairing signature in strongly attractive
  {Fermi} atomic gases.
\newblock {\em Physical Review B}, 76:085120, Aug 2007.

\bibitem{Liao_Nature_2010}
Y.~Liao, A.~S.~C. Rittner, T.~Paprotta, W.~Li, G.~B. Partridge, R.~G. Hulet,
  S.~K. Baur, and E.~J. Mueller.
\newblock Spin-imbalance in a one-dimensional {Fermi} gas.
\newblock {\em Nature}, 467:567, 2010.

\bibitem{Guan_PRA_2011}
X.-W. Guan and T.-L. Ho.
\newblock Quantum criticality of a one-dimensional attractive {Fermi} gas.
\newblock {\em Physical Review A}, 84(2):023616, 2011.

\bibitem{Fisher_PRB_1989}
M.~P.~A. Fisher, P.~B. Weichman, G.~Grinstein, and D.~S. Fisher.
\newblock Boson localization and the superfluid-insulator transition.
\newblock {\em Physical Review B}, 40(1):546--570, 1989.

\bibitem{Jiang_CPB_2015}
Y.-Z. Jiang, Y.-Y. Chen, and X.-W. Guan.
\newblock Understanding many-body physics in one dimension from the
  {Lieb–Liniger} model.
\newblock {\em Chinese Physics B}, 24(5):050311, 2015.

\bibitem{Cheng_PRB_2018}
S.~Cheng, Y.-C. Yu, M.~T. Batchelor, and X.-W. Guan.
\newblock {Fulde-Ferrell-Larkin-Ovchinnikov} correlation and free fluids in the
  one-dimensional attractive {Hubbard} model.
\newblock {\em Physical Review B}, 97(12):121111, 2018.
\newblock PRB.

\bibitem{Sachdev_2001}
S.~Sachdev.
\newblock {\em {Quantum Phase Transitions}}.
\newblock Cambridge University Press, Cambridge, 2001.

\bibitem{Lee_PRB_2017}
C.~H. Lee and C.~K. Gan.
\newblock Anharmonic interatomic force constants and thermal conductivity from
  {Gr\"uneisen} parameters: an application to graphene.
\newblock {\em Physical Review B}, 96(3), 2017.

\bibitem{Watanabe_JPSJ_2018}
S.~Watanabe and K.~Miyake.
\newblock Gr\"uneisen parameter and thermal expansion by the self-consistent
  renormalization theory of spin fluctuations.
\newblock {\em Journal of the Physical Society of Japan}, 87(3):034712, 2018.

\bibitem{Pethick_Cambridge_2008}
H.~Pethick, C. J.~Smith.
\newblock {\em {Bose-Einstein Condensation in Dilute Gases}}.
\newblock Cambridge University Press, Cambridge, 2 edition, 2008.

\bibitem{Yang_JMP_1969}
C.~N. Yang and C.~P. Yang.
\newblock Thermodynamics of a one-dimensional system of {Bosons} with repulsive
  delta-function interaction.
\newblock {\em Journal of Mathematical Physics}, 10(7):1115--1122, 1969.

\end{thebibliography}

\newpage

\ifNOSUP\end{document}